\documentclass[aps,prx,showpacs,floatfix,twocolumn,superscriptaddress,longbibliography,linenumbers]{revtex4-2}

\usepackage{amsmath}
\usepackage{amssymb}
\usepackage{amstext}
\usepackage{amsopn}
\usepackage{amsfonts}
\usepackage{amsxtra}
\usepackage[english]{babel}
\usepackage{graphicx}
\usepackage{float}
\usepackage{subfigure}
\usepackage{bm}
\usepackage{multirow}
\usepackage{mathrsfs}
\usepackage{mathtools}
\usepackage{dcolumn}
\usepackage{upgreek}

\usepackage[colorlinks,linkcolor=blue,urlcolor=blue,citecolor=blue]{hyperref}
\usepackage{color}
\usepackage{url}
\usepackage{breakurl}
\usepackage{relsize} 
\newcommand{\physunit}[1]{\ensuremath{\,\mathrm{#1}}}

\begin{document}
\nolinenumbers
\title{Revealing Hidden Inversion Symmetry Breaking in ZrTe$_5$ via Phonon-Assisted Heterodyne Amplification}

\author{S. J. Li}
\affiliation{Beijing National Laboratory for Condensed Matter Physics, Institute of Physics, Chinese Academy of Sciences, Beijing 100190, China}
\affiliation{School of Physical Sciences, University of Chinese Academy of Sciences, Beijing 100049, China}
\author{J. H. Huang}
\affiliation{International Center for Quantum Materials, School of Physics, Peking University, Beijing 100871, China}
\author{H. Y. Wu}
\affiliation{Beijing National Laboratory for Condensed Matter Physics, Institute of Physics, Chinese Academy of Sciences, Beijing 100190, China}
\affiliation{School of Physical Sciences, University of Chinese Academy of Sciences, Beijing 100049, China}
\author{S. P. Zheng}
\affiliation{Beijing National Laboratory for Condensed Matter Physics, Institute of Physics, Chinese Academy of Sciences, Beijing 100190, China}
\author{C. J. Kong}
\affiliation{Beijing National Laboratory for Condensed Matter Physics, Institute of Physics, Chinese Academy of Sciences, Beijing 100190, China}
\author{B. Xu}
\affiliation{Beijing National Laboratory for Condensed Matter Physics, Institute of Physics, Chinese Academy of Sciences, Beijing 100190, China}
\affiliation{School of Physical Sciences, University of Chinese Academy of Sciences, Beijing 100049, China}
\author{H. Wang}
\affiliation{International Center for Quantum Materials, School of Physics, Peking University, Beijing 100871, China}
\author{T. Dong}
\affiliation{Tsung-Dao Lee Institute, Shanghai Jiao Tong University, Shanghai 201210, China}
\affiliation{Zhangjiang Institute for Advanced Study, Shanghai Jiao Tong University, Shanghai 201210, China}
\author{L. Yue}
\author{D. Wu}
\affiliation{Beijing Academy of Quantum Information Sciences, Beijing 100913, China}
\author{Y. Wan}
\affiliation{Beijing National Laboratory for Condensed Matter Physics, Institute of Physics, Chinese Academy of Sciences, Beijing 100190, China}
\affiliation{School of Physical Sciences, University of Chinese Academy of Sciences, Beijing 100049, China}
\author{Z. L. Li}
\email{lizhilin@iphy.ac.cn}
\affiliation{Beijing National Laboratory for Condensed Matter Physics, Institute of Physics, Chinese Academy of Sciences, Beijing 100190, China}
\affiliation{School of Physical Sciences, University of Chinese Academy of Sciences, Beijing 100049, China}
\author{X. B. Wang}
\email{xinbowang@iphy.ac.cn}
\affiliation{Beijing National Laboratory for Condensed Matter Physics, Institute of Physics, Chinese Academy of Sciences, Beijing 100190, China}
\affiliation{School of Physical Sciences, University of Chinese Academy of Sciences, Beijing 100049, China}
\author{S. J. Zhang}
\email{sjzh@iphy.ac.cn}
\affiliation{Beijing National Laboratory for Condensed Matter Physics, Institute of Physics, Chinese Academy of Sciences, Beijing 100190, China}
\author{N. L. Wang}
\affiliation{Tsung-Dao Lee Institute, Shanghai Jiao Tong University, Shanghai 201210, China}
\affiliation{International Center for Quantum Materials, School of Physics, Peking University, Beijing 100871, China}
\affiliation{Beijing Academy of Quantum Information Sciences, Beijing 100913, China}
\affiliation{Zhangjiang Institute for Advanced Study, Shanghai Jiao Tong University, Shanghai 201210, China}
\author{Y. T. Li}
\affiliation{Beijing National Laboratory for Condensed Matter Physics, Institute of Physics, Chinese Academy of Sciences, Beijing 100190, China}
\affiliation{School of Physical Sciences, University of Chinese Academy of Sciences, Beijing 100049, China}
\affiliation{Attosecond Science Center, Songshan Lake Materials Laboratory, Dongguan, Guangdong 523808, China}

\begin{abstract}
ZrTe$_5$ is a sensitive topological material where small perturbations can alter its electronic structure. Its equilibrium crystal structure has been widely regarded as centrosymmetric, while recent experiments have raised the possibility of inversion-symmetry breaking. Here we probe this hidden symmetry lowering using nonlinear optical spectroscopy. Although conventional second-harmonic generation does not resolve an equilibrium symmetry-breaking signal, terahertz-field-induced second-harmonic generation (TFISH) reveals it through phonon-assisted heterodyne amplification. A coherently driven infrared-active phonon acts as a local oscillator for the vanishingly weak second-order susceptibility $\chi^{(2)}$, converting an otherwise undetectable symmetry-breaking response into a phonon-frequency modulation of the TFISH signal. The field-linear scaling of this modulation demonstrates $\chi^{(2)}$ is an equilibrium susceptibility rather than a response induced by the THz field. Polarization- and temperature-dependent measurements identify a bulk polar distortion along the crystallographic $a$ axis that persists to room temperature, while the $c$ axis remains nonpolar. These results provide direct optical evidence for equilibrium inversion-symmetry breaking in bulk ZrTe$_5$ and establish a structural constraint for understanding its electronic and topological properties.
\end{abstract}

\maketitle
%
%
Symmetry lowering is a sensitive fingerprint of hidden phases in quantum materials~\cite{annurev:/content/journals/10.1146/annurev-conmatphys-031218-013712}. Even a minute loss of symmetry can activate otherwise forbidden responses, lift symmetry-protected degeneracies, and reshape the electronic topology~\cite{RevModPhys.88.035005}. ZrTe$_5$ is especially sensitive because its band structure lies near a topological phase boundary~\cite{weng_transition-metal_2014,xu_temperature-driven_2018,WOS:000481798400015}. Small variations in lattice parameters, strain, temperature, stoichiometry, or carrier density can modify its electronic state~\cite{WOS:000398169900001,zhang_electronic_2017,WOS:000393196300038,shahi_bipolar_2018,xu_temperature-driven_2018,WOS:000481798400015,vaswani_light-driven_2020,WOS:000608683600007}, leading to a long-standing debate over its topological ground state~\cite{chen_optical_2015,manzoni_evidence_2016,wu_evidence_2016,WOS:000469334800022,jiang_unraveling_2020,zhang_observation_2021}.

Most electronic-structure discussions of ZrTe$_5$ have used the centrosymmetric space group $\mathit{Cmcm}$ ($D_{2h}^{17}$) as the structural starting point~\cite{SKELTON19821,FJELLVAG198691}. Recent transport experiments and first-principles calculations, however, have raised the possibility of inversion-symmetry breaking in ZrTe$_5$, calling this long-standing structural assumption into question~\cite{wang_gigantic_2022,wang_non-centrosymmetric_2024,xing_rashba-splitting-induced_2024,wang_spontaneous_2024}. Yet the nature of the symmetry-broken state remains unsettled, with conflicting reports on its polarity, temperature range, and occurrence in bulk crystals versus thin films. Resolving the equilibrium symmetry of ZrTe$_5$ is therefore essential because inversion-symmetry breaking introduces new nonlinear and geometric responses and modifies the symmetry constraints of its electronic structure~\cite{RevModPhys.90.015001,PhysRevLett.115.216806,RN1728}.

\begin{figure*}[tb]
    \centering
    \includegraphics[width=2\columnwidth]{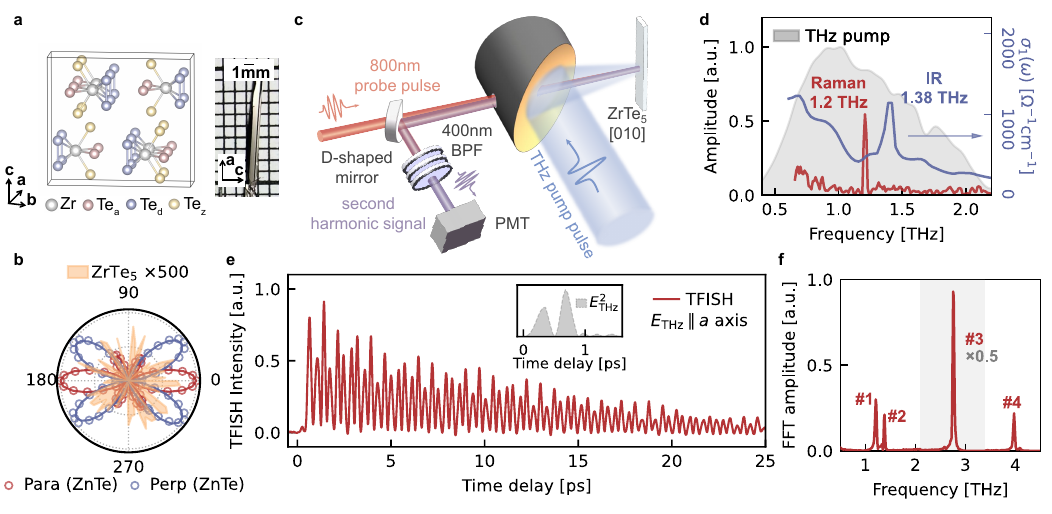}
    \caption{
    (a) Crystal structure of ZrTe$_5$. The strong anisotropy produces a needle-like morphology.
    (b) Rotational-anisotropy SHG pattern of equilibrium ZrTe$_5$ (yellow shading). No signal is resolved even after multiplying the trace by 500. ZnTe patterns measured under identical conditions are shown for comparison.
    (c) TFISH setup. BPF: band-pass filter; PMT: photomultiplier tube.
    (d) Frequency-domain characterization. The gray shading marks the THz pump spectrum. Optical conductivity (purple) reveals an infrared-active phonon, while optical-pump--optical-probe measurement (red) reveals a Raman-active phonon.
    (e) Time-domain TFISH signal at 5\physunit{K} under a 1\physunit{MV\,cm^{-1}} peak THz field. Inset: THz pump waveform.
    (f) Fourier spectrum of the TFISH signal, with Pk.~\#1--\#4 at 1.2, 1.38, 2.76, and 3.96\physunit{THz}. Pk.~\#3 is scaled by 0.5 for clarity.
    }
    \label{fig:fig1}
\end{figure*}

Second-order nonlinear optics provides a direct, contact-free probe of inversion symmetry. Within the electric-dipole approximation, the bulk second-order susceptibility $\chi^{(2)}$ is forbidden in centrosymmetric crystals and becomes allowed only when inversion symmetry is broken~\cite{shen2023second}, making second-harmonic generation (SHG) sensitive to subtle symmetry lowering~\cite{ZHAO2018207}. In ZrTe$_5$, however, the long-standing success of the $\mathit{Cmcm}$ structure suggests that any inversion-symmetry breaking, if present, is extremely weak. The associated inversion-odd distortion may therefore generate a minute electric-dipole $\chi^{(2)}$, whose static SHG intensity can fall below the detection threshold.

We therefore use terahertz-field-induced second-harmonic generation (TFISH) to reveal hidden equilibrium polarity in ZrTe$_5$ through phonon-assisted heterodyne amplification. In this approach, an intense THz pulse resonantly drives an infrared-active phonon, and the resulting modulation of the nonlinear susceptibility acts as a dynamic local oscillator for an otherwise undetectable equilibrium $\chi^{(2)}$. This converts the weak symmetry-breaking susceptibility into a pronounced phonon-frequency modulation of the TFISH signal, providing the key signature of a finite $\chi^{(2)}$ in ZrTe$_5$. The field scaling of this modulation distinguishes a pre-existing polar background from a THz-induced polarization, while polarization-resolved and temperature-dependent measurements determine the polar axis and its robustness. We identify a bulk equilibrium polar distortion along the $a$ axis that remains robust over the measured temperature range. These results establish inversion-symmetry breaking as a previously overlooked structural variable in ZrTe$_5$ and provide a basis for understanding its nonlinear optical and electronic properties.

The nonlinear-optical measurements were performed on chemical vapor transport (CVT)-grown ZrTe$_5$ single crystals, which exhibit a resistivity peak near 136\physunit{K}, consistent with previous reports~\cite{doi:10.1143/JPSJ.49.839,JONES1982793}. ZrTe$_5$ is a layered van der Waals material composed of ZrTe$_6$ trigonal prisms forming ZrTe$_3$ chains along the $a$ axis, linked by zigzag Te chains into sheets stacked along $b$ [Figure~\ref{fig:fig1}(a)].

Figure~\ref{fig:fig1}(b) shows the rotational-anisotropy SHG response of ZrTe$_5$ at 5\physunit{K} under normal-incidence reflection geometry. No static SHG signal is resolved above the noise floor, even after multiplying the trace by 500. As a calibration, noncentrosymmetric ZnTe measured under identical conditions gives the expected patterns~\cite{PhysRevB.58.10494}. This comparison bounds the static second-order susceptibility of ZrTe$_5$ near 800\physunit{nm} to below $0.5\physunit{pm\,V^{-1}}$.

In TFISH, an intense THz pulse resonantly drives an infrared-active phonon at $\Omega_{\rm IR}$, which modulates the nonlinear susceptibility~\cite{li_terahertz_2019,von_hoegen_amplification_2022,RN1569,RN271,li_terahertz_2025} and generates 
\allowdisplaybreaks
\begin{equation}
\begin{aligned}
I_{\mathrm{TFISH}}
&\propto
\left[
\left(
\chi^{(2)}
+
\frac{\partial \chi^{(2)}}{\partial Q_{\mathrm{IR}}}Q_{\mathrm{IR}}
\right)E^2
\right]^2  \\
&=
\left[
\left(\chi^{(2)}\right)^2
+
2\chi^{(2)}
\frac{\partial \chi^{(2)}}{\partial Q_{\mathrm{IR}}}Q_{\mathrm{IR}}
+
\left(
\frac{\partial \chi^{(2)}}{\partial Q_{\mathrm{IR}}}Q_{\mathrm{IR}}
\right)^2
\right]E^4 .
\end{aligned}
\label{equ:equ1}
\end{equation}
Here $\left(\partial\chi^{(2)}/\partial Q_{\mathrm{IR}}\right)Q_{\mathrm{IR}}$ denotes the phonon-induced modulation of the nonlinear susceptibility, and $E$ denotes the 800-nm probe field. Four-wave-mixing contributions from Dirac fermions~\cite{RN466,RN465,gao_terahertz_2024} are neglected, as they are limited to the $\sim$1-ps THz pulse.

\begin{figure*}[t] 
    \centering 
    \begin{minipage}[c]{1.5\columnwidth} 
        \centering
        \includegraphics[width=\linewidth]{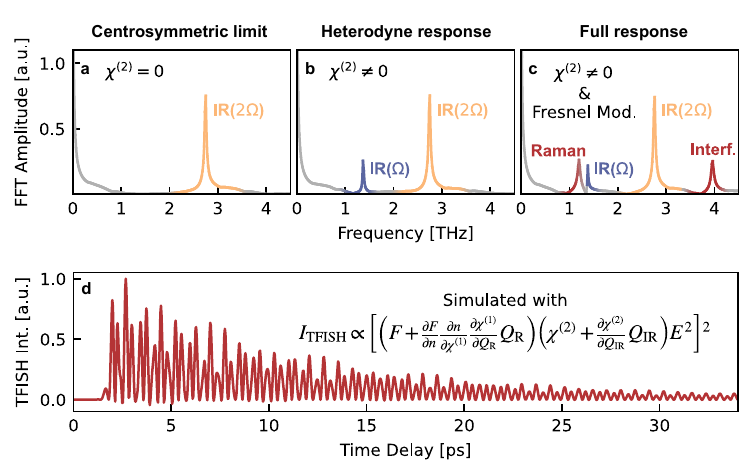} 
    \end{minipage}
    \hfill 
    \begin{minipage}[c]{0.55\columnwidth} 
        \caption{
    The simulated TFISH spectra show the stepwise buildup of spectral features as successive physical ingredients are included. (a) In the centrosymmetric limit, only the 2.76\physunit{THz} peak appears, corresponding to the second harmonic of the 1.38\physunit{THz} infrared-active phonon. (b) Broken inversion symmetry introduces a finite $\chi^{(2)}$, producing an additional peak at the infrared-phonon fundamental frequency, 1.38\physunit{THz}. (c,d) Fresnel-coefficient modulation by the Raman-active mode generates two further peaks at 1.2 and 3.96\physunit{THz}. The resulting spectra capture the main features of Figure~\ref{fig:fig1}(e,f).}
        \label{fig:fig2} 
    \end{minipage}
\end{figure*}

In Eq.~(\ref{equ:equ1}), the static term $\left(\chi^{(2)}\right)^2$ is removed by lock-in detection referenced to the THz-pump modulation. The quadratic phonon-induced term, $\left(\partial \chi^{(2)}/\partial Q_{\mathrm{IR}}\right)^2 Q_{\mathrm{IR}}^2$, produces only zero-frequency and $2\Omega_{\rm IR}$ components. The fundamental-frequency modulation at $\Omega_{\mathrm{IR}}$ instead arises from the heterodyne cross term $2\chi^{(2)}(\partial\chi^{(2)}/\partial Q_{\mathrm{IR}})Q_{\mathrm{IR}}$, which is linear in $Q_{\mathrm{IR}}$ and exists only when the equilibrium $\chi^{(2)}$ is nonzero. In ZrTe$_5$, low carrier density and weak scattering allow long-lived coherent phonons, enabling clear detection of this cross term.

This mechanism can be viewed as phonon-assisted heterodyne amplification. The phonon-induced modulation $(\partial \chi^{(2)}/\partial Q_{\mathrm{IR}})Q_{\mathrm{IR}}$ acts as the local oscillator and mixes with $\chi^{(2)}$, lifting the hidden symmetry-breaking signal above the detection threshold. The sensitivity arises from the large coherent amplitude of the THz-driven infrared phonon.

We then apply this heterodyne strategy to ZrTe$_5$. The experimental geometry is shown in Figure~\ref{fig:fig1}(c), with details in the Supplementary Information (SI). The THz pulse spans approximately 0.4--2.2\physunit{THz} [Figure~\ref{fig:fig1}(d)]. Figure~\ref{fig:fig1}(e) shows the time-domain TFISH signal measured at 5\physunit{K} with the THz electric field polarized along the $a$ axis and reaching a peak amplitude of 1\physunit{MV\,cm^{-1}}. A pronounced long-lived oscillation is observed. Its Fourier transform [Figure~\ref{fig:fig1}(f)] reveals four distinct sharp peaks at 1.2, 1.38, 2.76, and 3.96\physunit{THz}, labeled Pk.~\#1--\#4. From the full width at half maximum of these peaks, the lifetime of the oscillatory response is about 10\physunit{ps}, much longer than the $\sim 1\physunit{ps}$ duration of the THz pulse.

Optical conductivity [Figure~\ref{fig:fig1}(d)] and density-functional theory (DFT) calculations~\cite{aryal_robust_2022} show that a THz field polarized along the $a$ axis directly drives a single infrared-active phonon at 1.38\physunit{THz}. We therefore simulate the THz-driven phonon dynamics together with the corresponding TFISH responses, with details provided in the SI. In the centrosymmetric limit, $\chi^{(2)}=0$, the simulated TFISH response [Figure~\ref{fig:fig2}(a)] contains only a single peak at 2.76\physunit{THz}, corresponding to the second harmonic of the 1.38\physunit{THz} phonon, consistent with Pk.~\#3 and assigned as IR($2\Omega$). Thus, without inversion-symmetry breaking, the driven phonon generates only the $2\Omega_{\mathrm{IR}}$ response, not the fundamental $\Omega_{\mathrm{IR}}$ component.

Introducing a finite $\chi^{(2)}$ produces an additional peak at 1.38\physunit{THz} [Figure~\ref{fig:fig2}(b)]. This peak corresponds to the fundamental frequency of the driven infrared-active phonon and arises from the heterodyne cross term in Eq.~(\ref{equ:equ1}). We therefore assign Pk.~\#2 to the IR($\Omega$) component, which reflects a nonzero $\chi^{(2)}$ and serves as the primary TFISH signature of inversion-symmetry breaking.

\begin{figure}[!t]
    \centering
    \includegraphics[width=1\linewidth]{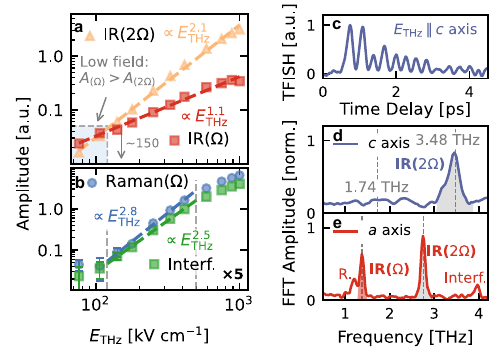}
    \caption{
(a) and (b) THz-field dependence of Pk.~\#1--\#4, assigned to Raman($\Omega$), IR($\Omega$), IR($2\Omega$), and interference (Interf.) components. The shaded region in (a) marks the regime where the amplitude of IR($\Omega$) is even stronger than that of IR($2\Omega$). Dots denote peak amplitudes; dashed lines are fits to the field dependences. Most error bars are smaller than the symbols.
(c) Time-domain TFISH signal at 5\physunit{K} with the THz polarization along the $c$ axis, under a 150\physunit{kV\,cm^{-1}} peak THz field.
(d) Fourier spectrum of (c).
(e) Fourier spectrum for THz excitation along the $a$ axis, under a comparable peak THz field of 150\physunit{kV\,cm^{-1}}. The IR($\Omega$) component is clearly observed for $a$-axis excitation in (e), in contrast to the absence of the corresponding fundamental-frequency component for $c$-axis excitation in (d).}
    \label{fig:fig3}
\end{figure}

We next examine the THz-field dependence of the IR($\Omega$) component. Figure~\ref{fig:fig3}(a) compares the amplitudes of the IR($\Omega$) and IR($2\Omega$) components for THz excitation along the $a$ axis. The IR($\Omega$) component is already visible at the lowest measured fields and shows no threshold-like onset. In the low-field regime, it is even stronger than the IR($2\Omega$) component, as marked by the shaded area. This behavior is inconsistent with a scenario in which the fundamental response appears only after the THz field induces a new polar state.

The field scaling provides more direct evidence. The IR($2\Omega$) component follows an approximately quadratic field dependence, consistent with the $Q_{\mathrm{IR}}^2$ term in Eq.~(\ref{equ:equ1}). By contrast, the IR($\Omega$) amplitude scales linearly with the THz electric-field amplitude $E_{\mathrm{THz}}$. Since $Q_{\mathrm{IR}}\propto E_{\mathrm{THz}}$ in the perturbative regime, this linear scaling shows that the $\chi^{(2)}$ factor in the heterodyne cross term is a constant over the measured THz-field range and already present before THz excitation. If $\chi^{(2)}$ were induced by the THz field, the fundamental response would vanish in the zero-field limit and acquire additional powers of $E_{\mathrm{THz}}$. The observed linear scaling therefore demonstrates that the nonlinear susceptibility is pre-existing rather than generated by THz excitation.

We now consider the remaining two peaks in Figure~\ref{fig:fig1}(f), which arise from the coupling between the Raman-active mode and the driven infrared phonon. The coherent Raman motion modulates the linear susceptibility $\boldsymbol{\chi}^{(1)}$~\cite{PhysRevB.45.768}, and hence the Fresnel factor. Mixing of this Raman-induced Fresnel modulation with the nonlinear response of the 1.38\physunit{THz} infrared-active phonon produces components at $\Omega_{\mathrm{R}}$ and $2\Omega_{\mathrm{IR}}+\Omega_{\mathrm{R}}$, corresponding to Pk.~\#1 and Pk.~\#4. A complete derivation of the Fresnel-modulation mechanism is provided in Section~S4.2 of the SI. Incorporating the 1.2\physunit{THz} Raman-active phonon into the simulation reproduces the main features of the experimental spectra [Figure~\ref{fig:fig2}(c,d)].

The field dependence of these two Raman-related peaks is shown in Figure~\ref{fig:fig3}(b). Both peaks scale approximately as $E_{\mathrm{THz}}^3$. Within the Fresnel-modulation picture, two powers of the THz field arise from the squared amplitude of the 1.38\physunit{THz} infrared-active phonon, leaving a linear field dependence for the 1.2\physunit{THz} Raman mode itself. Such field-linear excitation is not expected for a purely Raman-active phonon in a centrosymmetric lattice through conventional Raman pathways, such as impulsive stimulated Raman scattering (ISRS)~\cite{10.1063/1.449708} or sum-frequency excitation~\cite{maehrlein_terahertz_2017}. Displacive excitation of coherent phonons (DECP)~\cite{PhysRevB.45.768} is also unlikely given the low photon energy of the THz pulse, while other nonlinear mechanisms, including nonlinear phononics~\cite{RN403} and THz-field-induced tunneling~\cite{cheng_terahertz-induced_2025}, typically require substantially stronger fields. The residual linear field dependence therefore implies that the nominally Raman-active mode carries a finite effective mode charge, allowing it to couple directly to the THz electric field. This behavior provides independent support for a noncentrosymmetric equilibrium structure.

To identify the crystallographic direction of the equilibrium polarity, we compared the TFISH responses for THz excitation along the $a$ and $c$ axes. As shown in Figure~\ref{fig:fig3}(d), excitation along the $c$ axis produces only the $2\Omega_{\rm IR}$ component, while the fundamental-frequency response is clearly resolved for $a$-axis excitation under comparable fields. This anisotropy indicates that the equilibrium $\chi^{(2)}$ is associated with a polar distortion along the $a$ axis.

\begin{figure}[!t]
    \centering
    \includegraphics[width=1\linewidth]{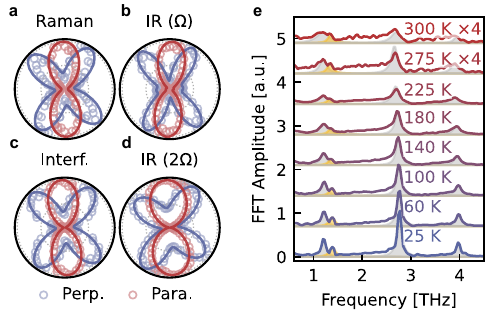}
    \caption{
    (a) Raman($\Omega$), (b) IR($\Omega$), (c) interference (Interf.), and (d) IR($2\Omega$) components. Red and blue symbols denote the integrated peak intensities measured in the parallel and perpendicular configurations, respectively. The solid lines are fits based on a model incorporating $\chi^{(2)}$, hyper-Raman, and Raman contributions. (e) Temperature-dependent TFISH spectra. The persistence of the IR($\Omega$) component (yellow shading) shows the inversion-symmetry breaking in ZrTe$_5$ survives up to room temperature. The spectra at 275 and 300\physunit{K} are multiplied by four for clarity.}
    \label{fig:fig4}
\end{figure}

To test whether this crystallographic anisotropy is consistent with a well-defined polar axis, we measured the rotational-anisotropy patterns of all four TFISH peaks in the parallel and perpendicular configurations, with the THz field fixed along the $a$ axis. These peaks reflect complementary nonlinear-response channels: IR($2\Omega$) is governed primarily by the hyper-Raman response of the driven infrared-active phonon; IR($\Omega$) arises from its interference with a finite $\chi^{(2)}$; and the Raman and interference peaks additionally involve Raman-induced Fresnel modulation. The fitting procedure is described in the SI. As shown in Figure~\ref{fig:fig4}(a)--(d), a model incorporating these contributions consistently reproduces the angular dependences of all four peaks in both detection geometries when the principal polar axis is aligned along $a$. Small rotations of the measured lobes do not alter this assignment and likely originate from residual optical aberrations, beam walking, or anisotropic Fresnel factors in the low-symmetry crystal~\cite{ZHAO2018207}. The IR($\Omega$) peak remains detectable up to room temperature [Figure~\ref{fig:fig4}(e)], showing that the polar distortion persists over the full temperature range studied. 

Taken together, the fundamental-frequency TFISH response, its field-linear scaling, the pronounced $a$-$c$ anisotropy, and its persistence to room temperature establish a hidden equilibrium polarity along the crystallographic $a$ axis in ZrTe$_5$.

A remaining question is whether the pre-existing $\chi^{(2)}$ originates predominantly from the bulk or from the crystal surface. A surface contribution associated with topological surface states can in principle generate a second-order response even when the bulk remains centrosymmetric~\cite{bowlan_probing_2017,melnikov_coherent_2018,melnikov_phonon-driven_2020}. For the CVT-grown ZrTe$_5$ crystals studied here, however, previous scanning tunneling microscopy~\cite{wu_evidence_2016} and angle-resolved photoemission spectroscopy~\cite{li_experimental_2016,zhang_electronic_2017,zhang_observation_2021} indicate that the $ac$ surface does not host topological surface states, making this origin unlikely. A generic surface-induced $\chi^{(2)}$ remains symmetry allowed, but it does not naturally explain the pronounced crystallographic anisotropy observed here. The $a$- and $c$-axis measurements were performed on the same cleavage surface, yet a robust fundamental-frequency response is observed only for $a$-axis excitation. Together with the normal-incidence geometry and polarization-resolved analysis in the SI, these observations favor a bulk polar distortion rather than a generic surface response.

The microscopic origin of the equilibrium polarity remains to be determined. Possible mechanisms include a collective lattice distortion, such as interlayer sliding predicted in ZrTe$_5$ films~\cite{wang_spontaneous_2024}, and defect-induced symmetry breaking in CVT-grown crystals~\cite{shahi_bipolar_2018}. Our measurements, however, favor a crystallographically correlated distortion over randomly oriented defect dipoles. The IR($\Omega$) response is tied to the 1.38\physunit{THz} infrared-active phonon, whose center frequency remains unchanged with THz field strength within our experimental resolution [Figure~S6 in the SI], indicating coupling to a well-defined lattice coordinate rather than field-induced defect rearrangement~\cite{cheng_terahertz-driven_2023}. Its field-linear scaling further shows that the phonon probes a pre-existing $\chi^{(2)}$ rather than creating the polar state~\cite{yang_prl_2025}. Moreover, the response is confined to the crystallographic $a$ direction, persists to room temperature, and remains reproducible after thermal cycling, supporting a robust equilibrium polar distortion with a well-defined orientation. An important feature of ZrTe$_5$ is that the equilibrium inversion breaking revealed here is extremely weak. The absence of a detectable static SHG response, together with the clear phonon-assisted TFISH modulation, indicates that the associated electric-dipole $\chi^{(2)}$ is much smaller than that of conventional noncentrosymmetric materials. This weak symmetry lowering places bulk ZrTe$_5$ close to the centrosymmetric limit.

Our nonlinear optical results complement previous transport evidence for inversion-symmetry breaking in ZrTe$_5$, including magnetochiral anisotropy, Rashba-related signatures, and the nonlinear Hall effect~\cite{wang_gigantic_2022,wang_non-centrosymmetric_2024,xing_rashba-splitting-induced_2024}. These measurements reveal important electronic consequences of inversion-asymmetric states but do not directly determine the underlying structural symmetry. In contrast, TFISH directly probes the lattice-enabled second-order susceptibility, identifies the polar axis through polarization-resolved measurements, and remains detectable up to room temperature without requiring a specific carrier-density regime. Our results therefore provide a structural-symmetry benchmark for interpreting electronic signatures associated with inversion-symmetry breaking in ZrTe$_5$.

The significance of this result lies in establishing equilibrium polarity as a previously overlooked structural variable in bulk ZrTe$_5$. Because ZrTe$_5$ lies close to a topological phase boundary, even weak inversion-symmetry lowering can modify the symmetry constraints on its electronic structure and influence the accessible topological phases~\cite{aryal_robust_2022,PhysRevB.103.075105,zhou_all-optical_2026}. Equilibrium polarity should therefore be considered alongside lattice parameters, strain, stoichiometry, temperature, and carrier density when describing the electronic and topological properties of ZrTe$_5$.

In summary, we have identified a weak polar distortion in bulk ZrTe$_5$ that persists to room temperature while producing a static SHG response below our detection limit. The evidence comes not from a marginal intensity enhancement, but from a coherent, symmetry-selective heterodyne response. A THz-driven infrared phonon makes a pre-existing $\chi^{(2)}$ visible as a sharp fundamental-frequency oscillation. Its field scaling, polarization-resolved response, persistence to room temperature, and strong a–c anisotropy identify equilibrium bulk inversion-symmetry breaking along the crystallographic $a$ axis. The combination of a null static-SHG response and a clear phonon-assisted TFISH modulation establishes weak but finite equilibrium polarity in ZrTe$_5$, placing a concrete structural constraint on its topological description. More broadly, this work provides a route to detect hidden polar phases in quantum materials where the relevant distortion is too weak for standard nonlinear optics. It also provides criteria for distinguishing pre-existing weak inversion-symmetry breaking from genuinely THz-induced polar responses, a distinction essential for probing and steering emergent quantum states through coherent lattice motion.


\begin{center}
\small{\textbf{ACKNOWLEDGMENTS}}
\end{center}
This work was supported by the National Key Research and Development Program of China (Nos.~2024YFA1408700 and 2023YFA1607400) and the National Natural Science Foundation of China (Nos.~12488201, U24A2016, 12304184, 12574164, and 12574349). This work was carried out at the Synergetic Extreme Condition User Facility (SECUF, https://cstr.cn/31123.02.SECUF). Z.L. was supported by the Youth Innovation Promotion Association of CAS (No.~2021008). 

\bibliography{ref}

\end{document}